\documentstyle[preprint,aps,prb,epsf]{revtex}   

\begin{document}

\title{{\bf Kinetic non-optimality and vibrational stability of proteins}}

\author{{\bf Marek Cieplak$^*$ and Trinh Xuan Hoang}}

\address{Institute of Physics, Polish Academy of Sciences,
Al. Lotnikow 32/46, 02-668 Warsaw, Poland }

\maketitle

\vskip 40pt
\noindent
$^*$Correspondence to: \\
Marek Cieplak,\\
Physics Department, \\
Rutgers University, \\
Piscataway NJ-08854, USA. \\
Tel:  732-445-1440\\
Fax:  732-445-4343\\
E-mail: cieplak@physics.rutgers.edu

\vskip 40pt
\noindent
Grant sponsor: KBN (Poland); Grant number: 2P03B-146-18.


\newpage
\begin{abstract}
{\bf Scaling of folding times in Go models of proteins 
and of decoy structures with the
Lennard-Jones potentials in the native contacts reveal 
power law trends when studied under optimal 
folding conditions. The power law exponent depends
on the type of native geometry. Its value indicates lack
of kinetic optimality in the model proteins.
In proteins, mechanical and thermodynamic stabilities are
correlated.}
\end{abstract}

\vspace*{0.5cm}

\section*{INTRODUCTION}

Proteins are extraordinary heteropolymers.  They fold to their
native states much faster than what a blind combinatorics would
predict \cite{Levinthal} since a folding funnel in the energy
landscape is formed.\cite{Leopold,Bryngelson1,DillChan} 
Proteins are believed to
have high designabilities,\cite{Tang} to be stable against
mutations,\cite{Vendruscolo,BauerChan} 
and to have the highest densities
of states.\cite{Michel1} Furthermore, the $\alpha$ helix secondary motifs
have been shown  theoretically to be the fastest folders among
chains of the same number, $N$, of 
aminoacids \cite{Michel} and be
the result of the geometrical optimization of compact
chains with maximum wiggle room. \cite{Maritan}
Experimental 
results \cite{Munoz,Plaxco,Ruczinski}
(see also a commentary by Chan \cite{Chan}) also point to the
accelerating role of the helices.
Biological evolution may have optimized functionality of
proteins, but can proteins be optimal kinetically?

Here, we consider folding times, $t_{fold}$, in Go models
\cite{Goabe,Hoang,Hoang1} of proteins and decoy structures and show that
though proteins fold to their native structures fast
they are not optimal folders.
This conclusion ties well with the protein engineering
experiments \cite{KimBaker,Fersht} which show that mutations in
wild type proteins may lead to significant increases in folding
rates and thus show no kinetic optimality of sequences.  Our
theoretical argument is based on relating universality classes in
the scaling of $t_{fold}$   to classes of native geometries.
This confirms  a decisive role of native geometry in determining
properties of proteins.\cite{Baker}  The scaling trends that we
observe are robust when studied at the temperature of the fastest
folding, $T_{min}$, but become obscure when studied at other
temperatures.

Another issue which we examine here regards the notion of protein
stability.  One definition of stability is thermodynamic -- it
assesses the role of non-native phase space valleys relative to
the native valley   by determining the probability of staying in
the native basin.  It is characterized by the folding
temperature, $T_f$, at which this probability crosses
$\frac{1}{2}$.\cite{Socci}  Another is mechanical: at what
temperature will the native conformation melt due to vibrations?
The mechanical definition  does not refer  to non-native valleys.
The two notions should correlate with each other if the native
valley dominates in the energy landscape. We show that this is
indeed what happens in model proteins.

\vspace*{0.5cm}

\section*{MODEL AND METHOD}

We first consider the problem of universality classes in the
scaling of folding properties.  There have been various
predictions about the nature of scaling of $t_{fold}$. A number
of theories suggested a power law dependence of barrier heights
on $N$ and thus an exponential law for 
$t_{fold}$.\cite{Takada,Finkelstein,Wolynes2}  
Thirumalai,\cite{Thirumalai3} however, has argued in favor of a power law
for $t_{fold}$, 
\begin{equation}
t_{fold} \sim N^{\lambda} \;\;,
\end{equation}
where $\lambda$ is estimated to be between 3.8
and 4.2 for simple two-state folders.  A heuristic model
\cite{Camacho}  leads to $\lambda =3$.  Numerical studies of
$t_{fold}$ in various lattice models
\cite{Shakhnovich,Zhdanov,Cieplak} have supported the power law
behavior and indicated dependence of $\lambda$ on specifics of
the model, dimensionality and temperature, $T$.  For designed
sequences in three dimensions, $\lambda$ has been found to be in
the Thirumalai range \cite{Shakhnovich} whereas for Go models it
has been found to be of order 3.\cite{Shakhnovich,Cieplak} In
these studies, $t_{fold}$ is defined as the first passage time.

Here, we extend the scaling studies to off-lattice Go models
\cite{Goabe} and consider chains of of beads separated by $d_0
\approx 3.8 \AA$ -- a typical length of the peptide bond.  The Go
Hamiltonian is defined through a native conformation of a
sequence since it assigns relevant interaction energies only to
the native contacts.  Despite this simplification, the  Go models
may behave more realistically than atomistic models.\cite{Stakada}

It should be noted that the Go models are so minimal that they
disregard an explicit amino acidic definition of a protein
and variablity of the volume taken by individual side chains.
Natural proteins appear to fold by locking its segments together in
an unfrustrated way. Adding attraction to the non-native
contacts in the bead-spring model might seem to be making the
model more realistic but, in fact, it leads to spurious entanglements
during folding. In this sense, the Go model repairs some of its
shortcomings by a mutual cancellation of its ills and focuses
on the effects related to the native structure. This 
focus is justified by experimental indications that
the native structure itself is central to folding. \cite{Alm,Ruczinski}
On the other hand, the target oriented aspects of such theoretical
modeling are hard to justify on a fundamental level. \cite{Kaya}
The nature of the Go model allows one to study the role of the
native structure in kinetics but it does not allow to address
the role of the sequential order. Determining sequence based,
as opposed to structure based, classes of kinetic universality
would be much more interesting but, clearly,
also much more challenging.

We employ Lennard-Jones (LJ) potentials
$V_{ij} = 4\epsilon \left[ \left( \frac{\sigma_{ij}}{r_{ij}}
\right)^{12}-\left(\frac{\sigma_{ij}}{r_{ij}}\right)^6\right] $
for the native contact interactions between monomers $i$ and $j$,
in a distance of $r_{ij}$ apart. The non-native interactions
are described by 
repulsive soft core potentials 
that provide excluded volume and prevent entanglements.
Our approach is presented in details in Ref.\cite{Hoang,Hoang1}
where several secondary structures and three model proteins have been 
analyzed. Such models were also studied in Ref.\cite{Li,Klimov2} The distances
between successive beads are controlled by an anharmonic potential. 
The length parameters $\sigma _{ij}$ are
selected so that the minimum of $V_{ij}$ corresponds to the
geometry found in the target structure and the contacts are said 
to be formed when $i$ and $j$ are not consecutive along the chain and $r_{ij}$
is less than $d_{nat}$, where $d_{nat}$ is 
$7.5 \AA$.

There are other variants of the off-lattice Go models: 
Zhou and Karplus \cite{Zhou} and Dokholyan et al. \cite{Stanley}
have considered models with a square well potential. 
Clementi et al. \cite{Nymeyer} have studied the 12-10
power law potentials.
It is not clear which effective potential is the best and our
choice is LJ.

The dynamics of the system are described by the Langevin equation
$ m\ddot{{\bf r}} = -\gamma \dot{{\bf r}} + F_c + \Gamma  $
where ${\bf r}$ is a position of a monomer, $m$ is its 
mass and $F_c$ is the force derived from the Hamiltonian.
$\gamma$ is a friction
coefficient and $\Gamma$ is the random force such that
$\left<\Gamma(0)\Gamma(t)\right> = 2\gamma k_B T \delta(t) \;$,
where $k_B$ is the Boltzmann constant, $t$ is time and
$\delta(t)$ is the Dirac delta function.  Both the friction and
the random force represent the effects of the solvent and they
control $T$.  The equations are solved using the fifth order
predictor-corrector scheme.

In the following, $T$ is measured in the units of $\epsilon/k_B$
and $t$ is measured in units of the oscillatory period $\tau$.
At low values of friction, $\tau$ is equal to
$\sqrt{ma^2/\epsilon}$, where $a$ is a van der Waals radius of
the amino acid residues. The value of $a$ is chosen as $5\AA$
which is roughly equal to $\left<\sigma_{ij}\right>$ in our model
proteins.  The simulations are done with $\gamma=2 m\tau^{-1}$ --
a standard choice in studies of liquids. Higher values of
$\gamma$ have been argued to be more realistic.\cite{Veitshans}
We have shown \cite{Hoang} that $t_{fold}$ is linear in $\gamma$
and $T_{min}$ depends on $\gamma$ weakly.

The native conformation is defined through the locations of the
$\alpha$ carbons.  We have considered 21 single domain Protein
Data Bank (PDB) structures\cite{PDB} with $N$ ranging between 29
and 98.  9 of these structures belong to a set of proteins
considered by Plaxco et al. \cite{Plaxco} or are their close
homologies.
These are: the SH3 domain of 1efn (57), 2ptl (63), 
2ci2 (83-18=65; 18 are not resolved), 1csp (67), 1ubq (76),
1hdn (85), 2abd (86), 1ten (90), and 1aps (98), 
where the numbers in brackets indicate the corresponding values of $N$.
The additional 12 structures are:
1cti (29), 1cmr (31), 1ce4 (35), 1bba(36), 1erc (40), 
1crn (46), 7rxn (52), 5pti (58), 1tap(60), 1aho (64), 1ptx (64), 1erg (70).
These conformations were picked from the low-$N$ end of the size
distribution to allow for a reliable characterization. Our
studies of these structures indicate well defined overall trends
in $t_{fold}$  which are only weakly affected by an inclusion of
steric constraints.\cite{Veitshans}  Our results will be given here
only for models without such constraints. 

The results obtained for the PDB structures are compared to five
classes of decoy  conformations which differ in the way they fill
space and in their packing arrangements.  These classes form
statistical ensembles in which a given value of $N$ has multiple
realizations.  Four classes are defined in terms of shapes that
homopolymers arrive at under various cooling procedures.  The
non-consecutive beads in the homopolymers interact through the LJ
potential with $\sigma _{ij}= 5\AA$ which corresponds to a
typical  van der Waals radius of aminoacids. We discuss the
following classes (see Figure 1): 

\begin{description}

\item HC: conformations obtained through slow homopolymer cooling.
The procedure involves generating an open conformation,
assigning identical strengths to all inter-bead interactions, and then
slowly annealing. 
The resulting compact conformation serves as a
native structure in the Go-like Hamiltonian.

\item HQ: similar to HC but with a rapid quenching instead of annealing.
The procedure results in non-compact native structures which, however,
have many local contacts, as measured along the chain, and are thus
more closely related to $\alpha$-helices than to random heteropolymers. 

\item HA: similar to HC but the $\alpha$-helices of various lengths
(of order 15) are first built into the initial states
in a way which is consistent with the LJ couplings
and then preserved through the
annealing process by assigning ten times stronger couplings 
to the helical secondary structures.

\item HB: similar to HA but the helical segments are replaced by
$\beta$-sheet conformations.
The lengths of the $\beta$-strands are fixed at 8 monomers.

\item CL: compact native conformations generated on a grid as a
self-avoiding random walk within a compact box of lattice constant
equal to the length of a peptide bond and then stabilized by
appropriate Lennard-Jones interactions.

\end{description}

The folding properties
are studied as a function of $T$ and then presented here 
for $T$=$T_{min}$ and $T_f$.
at which probability, $P_0$, of being in the native basin is $\frac{1}{2}$.
$P_0$ is determined based on 10-15 long molecular dynamics trajectories
at equilibrium. The results are illustrated in Figure 2 for two
model proteins 1ubq and 1ce4.

The median folding times depend on $T$ in a U-shaped fashion and, 
generally, the bigger the $N$, the narrower the U.
The dependence of $t_{fold}$ for the Go models of 1ubq and 1ce4
Is shown at the bottomn of Figure 2.

The system is assumed to be in its native state if
all of its native contacts are established.
A  native contact between monomers $i$ and $j$ is said
to be established if $r_{ij}$
is shorter than 1.5$\sigma_{ij}$.

\vspace*{0.5cm}

\section*{RESULTS AND DISCUSSION}

\noindent
{\bf Kinetic Universality Classes}

Figure 3 shows 
the validity of the power law, eq. (1), for
$t_{fold}$ when determined   
under the most favorable kinetic conditions -- at $T_{min}$.
The exponent $\lambda$ sensitively depends on the 
geometrical class of the native structures 
(Table I). The case of HB is special since a
crossover between two effective values of $\lambda$ is observed
(the $\beta$-strand lengths of 8 impose a condition on $N$ above
which a characteristic $\beta$-sheet behavior can start to be seen).
The values of $\lambda$ range between 1.7 and 3.2. 
The smallest $\lambda$ 
corresponds to the HA 
and the largest to the HQ and long HB conformations. HC is intermediate.
Note that $\lambda$ for HA is smaller than 2 -- the 
value suggested by de Gennes \cite{Gennes} in his analysis
of the time scale for the coil to globule transition of a homopolymer.

The data points for PDB at $T_{min}$ are somewhat scattered --
there is no averaging over an ensemble --
but a well defined trend is visible. The 
exponent $\lambda$ is 
about $2.5 \pm 0.2$  which 
indicates that these structures
are not optimal kinetically. HA, short HB, and HC of the same $N$ fold
faster. PDB appears to be comparable to the grid conformations CL
(there is only a week dependence on the dimensionality in the 
off-lattice models -- when the grid structures are generated on the
square lattice, $\lambda$ becomes equal to $2.1 \pm 0.2$).

The existence of a 
trend in the scaling of $t_{fold}$ for the PDB
structures appears to be at odds with the analysis of experimental
data compiled by Plaxco et al.,\cite{Plaxco} 
and replotted here in Figure 4, which indicated
lack of any correlations with $N$. 
A flavor of this is already seen in Figure 3 which shows that one
sequence (1aps) has a $t_{fold}$ which is distant from the
trend. This sequence appears to be frustrated geometrically
and it has a very small experimental folding rate 
\cite{Plaxco} -- maybe it is just a poor folder.
However, the experimental data indicate a significantly larger
scatter of values than 
as seen in Figure 3.

There are three explanations of the discrepancy that we considered. 
First: the range
of the values of $N$ considered in Ref.\cite{Plaxco} is smaller
than 
studied here, which in itself emphasizes fluctuations.
However, in data  that were published later \cite{Ruczinski}
the range of $N$ was extended to about 150 and the correlations of
kinetics with $N$ remained weak so the limited range of the
values of $N$ is not a likely explanation of the discrepancy.
Second, it is only the simplified models, like the Go models, that
show trends in the kinetics of folding whereas any additional
complexities present in real systems may perturb such trends
beyond detection. This possibility could be studied in the future
by considering scaling in more complicated classes of models.
In particular, the role of the localization index of the
interactions should be elucidated in the context of scaling.
Third: the trends are derailed by the fact that
the experimental data are usually obtained at a fixed
temperature, typically, but not necessarily, at the room temperature.
Thus the data collection involved no kinetic optimization which would
require selecting the best $T$
for each protein individually. 

The role of this third possibility is illustrated at the
top of Figure 4 which shows the scaling of $t_{fold}$ at $T_f$.
The scatter is seen to be significantly larger than at $T_{min}$.
It is not as large as in the experimental data but 
it should be noticed, again, that our Go systems
are just very simple models of systems which are
quite complicated.
Another way to asses the relevance of the optimal selection of $T$
is shown in Figure 5 which reanalyzes the data of figures 3 and 4 so that
theoretically determined $t_{fold}$ is plotted against 
the experimentally measured folding time, $t_{exp}$.
The small number of available points makes it hard to place
a bet on the best trend. However, it is clear that the points
determined at $T_{min}$ exhibit significantly less scatter
than those calculated at $T_f$. This finding gives further support
for the idea that the lack of optimization in the temperature may
mask existence of any underlying trends.

Our findings on scaling of characteristic $T$'s can be summarized
as follows. For HC, HA and CL, $T_{min}$ grows with $N$
whereas $T_f$ is almost constant and somewhat lower than $T_{min}$.
The difference between $T_{min}$ and $T_f$ grows most slowly for
HA.
For PDB, $T_{min}$ does not seem to have a trend, within the range
of $N$ studied, and the values of $T_{min}$ are usually just above
the corresponding values of $T_f$. This indicates a borderline behavior
between excellent and poor folding characteristics, if the condition
for the latter is $T_f << T_{min}$.\cite{Socci,Henkel} 
This borderline behavior 
might characterize classes of proteins, especially of those which
have a short lifetime in a living cell but
this result may depend on the choice of the potentials.

\vspace*{0.5cm}

\noindent
{\bf Stability Against Vibrations}

We now discuss stability of the native state in proteins.  The
mechanical stability can be probed  through the phononic spectra
as in Ref.\cite{Hoang}. This is accomplished by determining the
frequency gap, $\omega _1$ in the low end of the frequency
spectrum. Another test is provided by studying root mean square
displacements around the native state and employing the Lindemann
criterion for melting.  We introduce the parameter
\begin{equation}
\delta _L (T)\;=\; 
\frac{1}{n} \sum _{i>j} ^{[{\rm NAT]}}
\frac{\sqrt{\left<r^2_{ij}\right> -
\left<r_{ij}\right>^2}}{r_{ij,\rm NAT}} \;,
\end{equation}
which is a variant of the parameter used by Takano et al.
\cite{Takano}  The summation is  over pairs of monomers ($n$ of
them) which form native contacts and their rms displacement is
compared to the native distances. The temperature, $T_L$, at
which $\delta _L$ crosses the Lindemann value of 0.1  is a
measure of mechanical stability.  Figure 6 shows that both
$\omega _1$ and $T_L$ show a correlation with $T_f$ which
suggests the predominance of the native valley in the energy
landscape. Note that $T_L$ is higher than $T_f$ which indicates
that the probability "leaks out" of the native state already when
the vibrations in the native valley are small.  Thus for good
folders, the notions of thermodynamic and mechanical stabilities
qualitatively coincide.

\vspace*{0.5cm}

\section*{CONCLUSIONS}

Our main results on scaling can be summarized as follows.
There are kinetic universality classes
among well folding sequences. These classes depend on the type
of geometry involved in the native state. Well defined scaling trends
can be established if folding is studied under optimal conditions.
Otherwise they  are hard to be seen, especially if the range of system
sizes is narrow.
The shapes of actual proteins in their native states are such
that the folding times scale with an exponent which is higher than
certain artificial classes of structures. This suggests the lack
of kinetic optimality of proteins.

Our results have been obtained within the Go model which focuses
on the role of the native state geometry. This level of
simplified description incorporates a long list of approximations
which somehow appear to compensate mutually. In effect, the Go 
systems are reasonable models of good folders and the simplifications
involved are precisely of the kind that allow for a statistical
analysis necessary to establish scaling properties. Working with
sequences described in a more sophisticated manner 
would add to the reality of description. However,
it would also necessitate dealing with statistical
ensembles of sequences defined by more parameters than just
the size and shape and currently that would be prohibitive numerically.
Our results should then be viewed as establishing first inroads
into understanding of the role of size in folding kinetics.

\section*{ACKNOWLEDGMENTS}
We appreciate discussions with J. R. Banavar, C. Johnson, H. Nymeyer, 
J. N. Onuchic, S. Plotkin, and D. Thirumalai 
which motivated parts of this research. This work was funded
by KBN.


\vbox{
\begin{table}[t]
\caption{The values of exponent $\lambda$ for the classes
of conformations studied.}
\vspace{10pt}
\begin{tabular}{cc}
Structure & $\lambda$ \\
\hline
HC   &  $2.2 \pm 0.1$ \\
HA   &  $1.7 \pm 0.1$ \\
HB   &  $0.9 \pm 0.1$, $3.2 \pm 0.1$ \\
HQ   &  $2.7 \pm 0.2$ \\
CL   &  $2.6 \pm 0.2$ \\
PDB  &  $2.5 \pm 0.2$ \\
\end{tabular}
\end{table}
}

\newpage
\centerline{FIGURE CAPTIONS}

\begin{description}
\item[Fig. 1.]
Examples of native conformations 
that were used in these studies. 
The folding data were generated based on 11 realizations of each
class of structures for each value of $N$, 
except for the case of HA when
5 realizations were 
sufficient.
\item[Fig. 2.] Equilibrium and kinetic properties of the Lennard-Jones-Go
model of two proteins, 1ce4 (solid lines) and 1ubq (dotted lines), as a
function of temperature. The top panel shows the probability of staying in the
native state and the bottom panel shows the median folding time
as determined based on 200 trajectories at each temperature.
\item[Fig. 3.]
Power law dependence of the median folding times
at $T_{min}$ on $N$ for the indicated classes of geometry
of the native conformations.
The proteins analyzed by Plaxco et al.\cite{Plaxco} are shown as 
open squares. Closed squares correspond to other proteins.
For CL, the times are multiplied by 10. 
\item[Fig. 4.]
Top: Values of $t_{fold}$ for the PDB structures as determined
at $T_{f}$. The open circles refer to the proteins studied by 
Plaxco et al.\cite{Plaxco} The 
line shows the scaling trend 
found at $T_{min}$.
Bottom: Experimentally determined folding times, $t_{exp}$, (inverses
of the folding rates) as compiled
by Plaxco et al.\cite{Plaxco}
\item[Fig. 5]
Folding times of the theoretical Go model  versus folding times
observed experimentally in proteins studied by Plaxco et al. \cite{Plaxco}.
The solid and open squares correspond to $T=T_{min}$ and $T=T_f$
respectively.
\item[Fig. 6.]
Bottom: $T_L$ as a function of $T_f$ for the PDB structures studied.
Top: The lowest non-zero phononic frequency of the same
structures plotted vs. $T_f$.
The broken lines indicate overall trends.
The inset illustrates the dependence of $\delta _L$ (eq. 2) on 
temperature, $T$, for  the Go model of crambin.
The horizontal line indicates the 10\% value of  $\delta _L$.
\end{description}

\begin{figure}
\epsfxsize=4.5in
\centerline{\epsffile{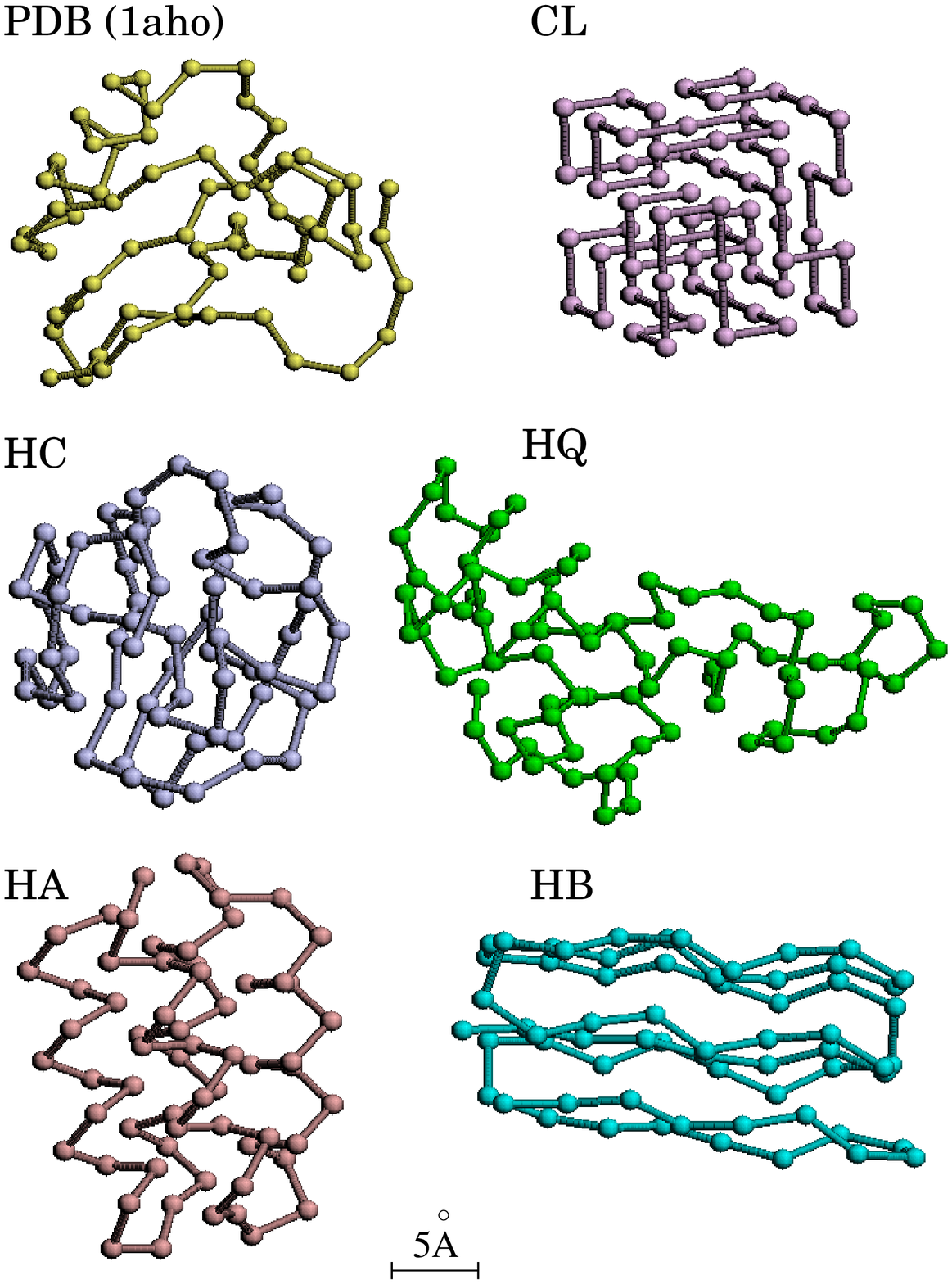}}
\caption{ }
\end{figure}

\begin{figure}
\epsfxsize=4.5in
\centerline{\epsffile{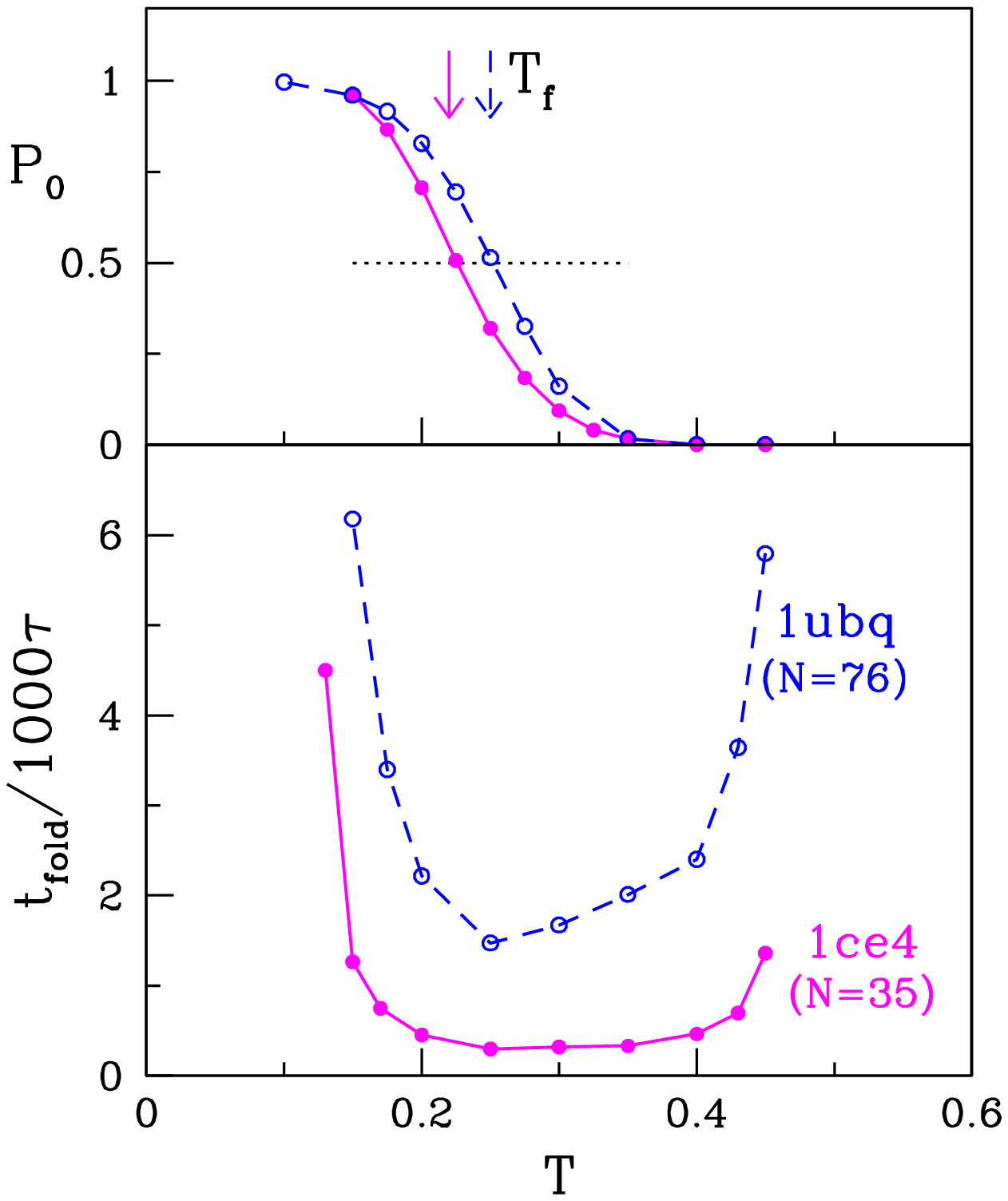}}
\caption{ }
\end{figure}

\begin{figure}
\epsfxsize=4.5in
\centerline{\epsffile{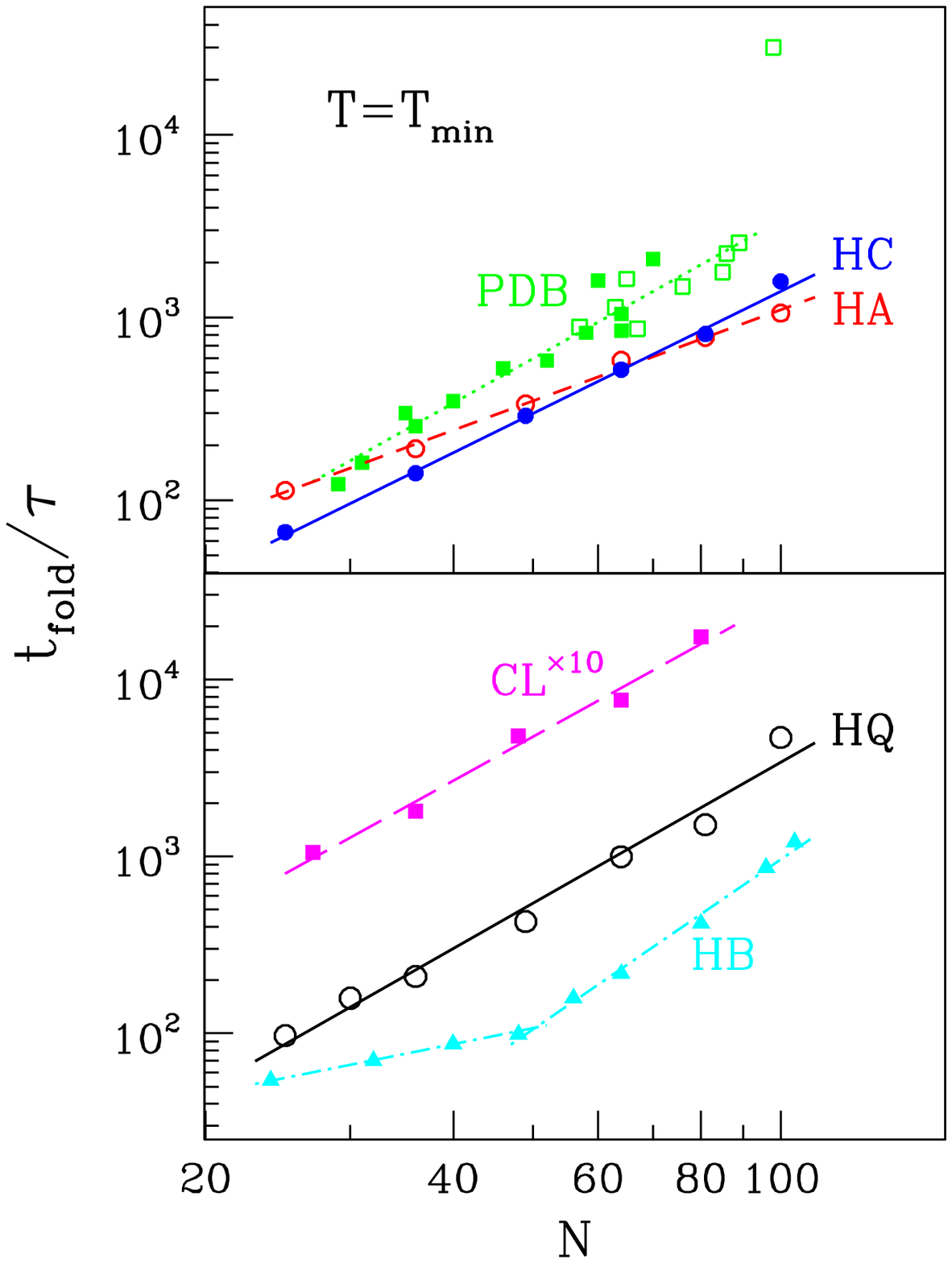}}
\caption{ }
\end{figure}

\begin{figure}
\epsfxsize=4.5in
\centerline{\epsffile{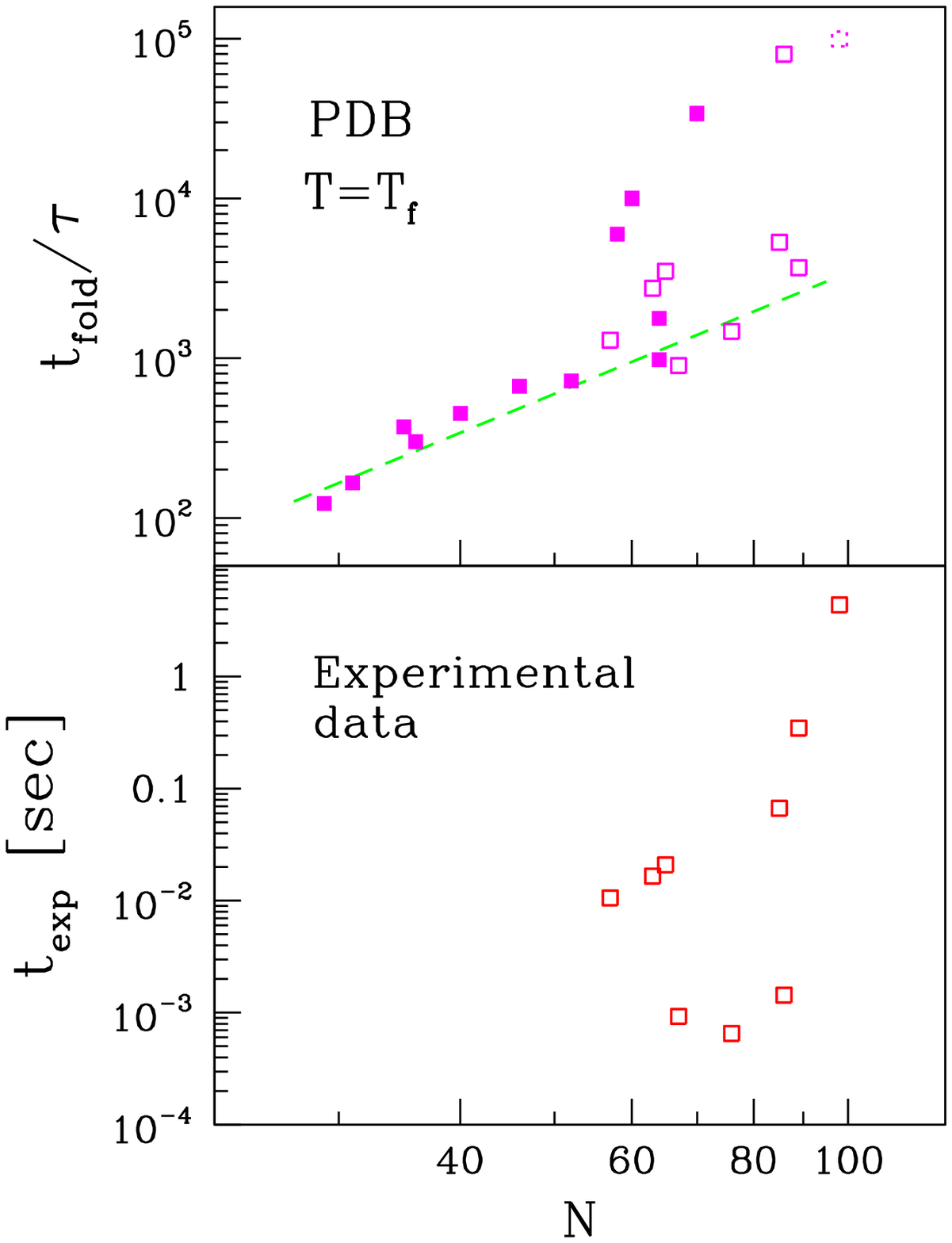}}
\caption{ }
\end{figure}

\begin{figure}
\epsfxsize=4.5in
\centerline{\epsffile{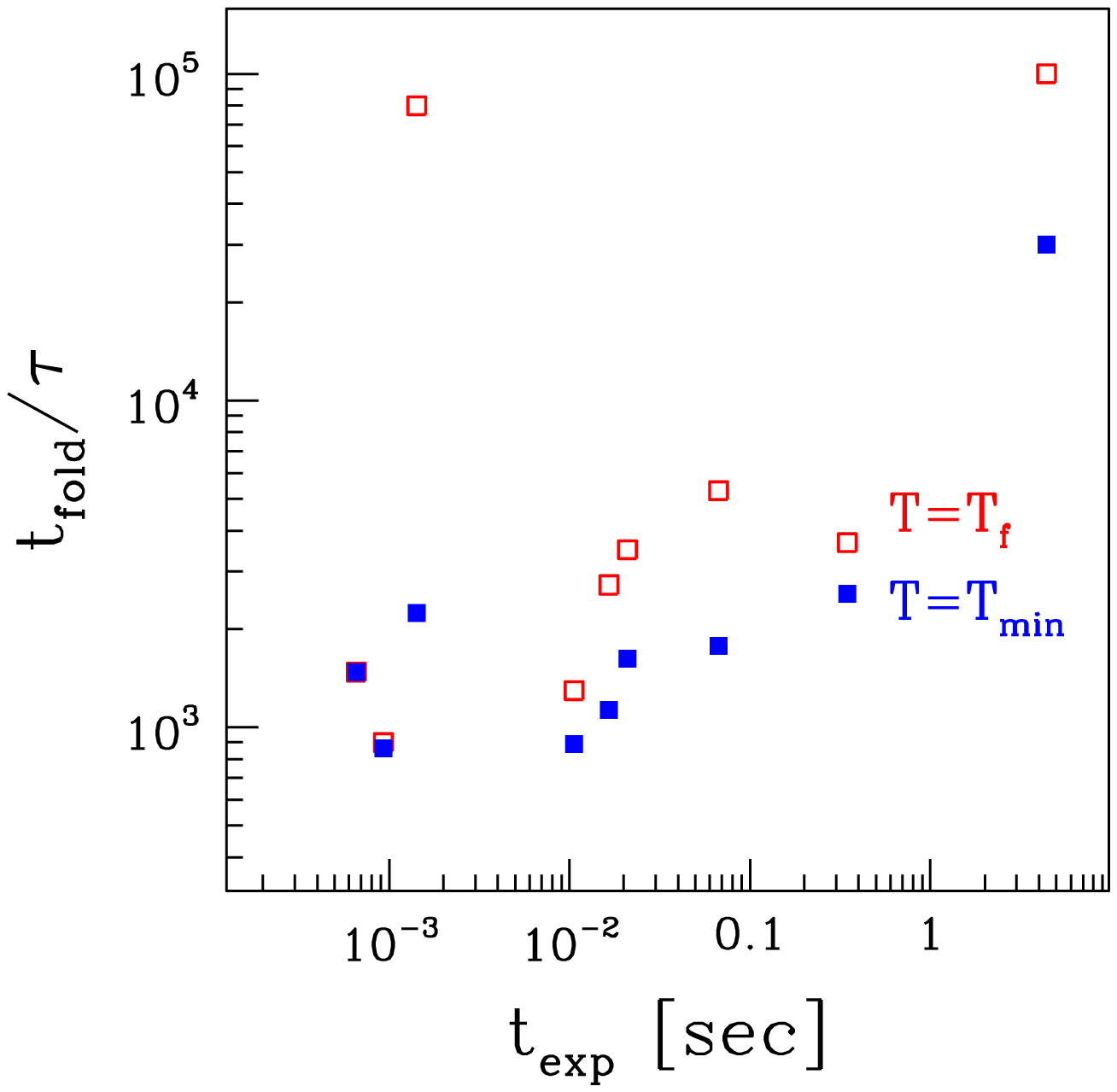}}
\caption{
}
\end{figure}

\begin{figure}
\epsfxsize=4.5in
\centerline{\epsffile{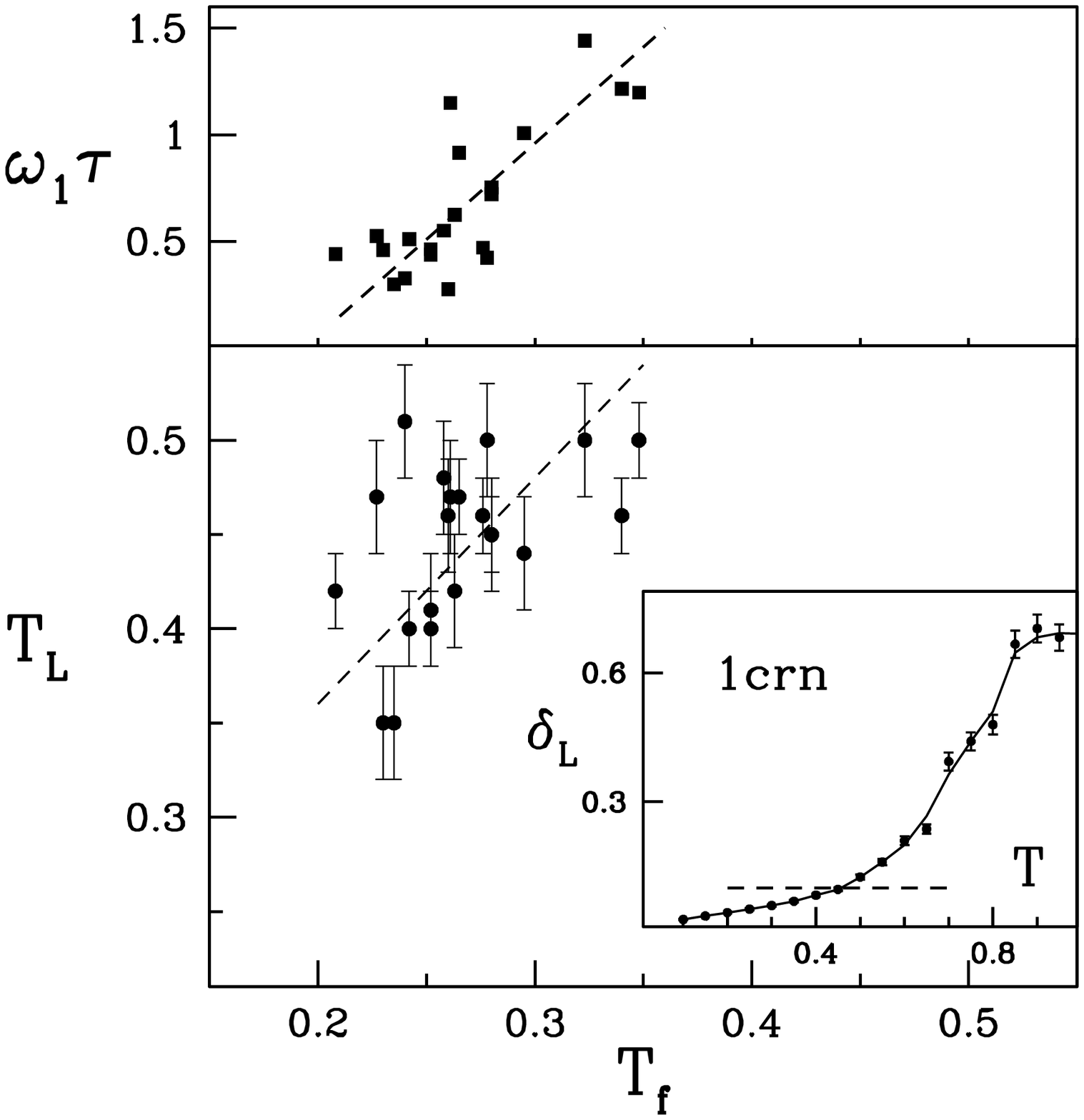}}
\caption{
}
\end{figure}

\begin{thebibliography}{99}

\bibitem{Levinthal}
Levinthal C. In: 
Debrunner P, editor.
Mossbauer Spectroscopy in Biological Systems,
Urbana: University of Illinois Press; 1969.


\bibitem{Leopold}
Leopold PE, Montal M, Onuchic JN.
Protein folding funnels: a kinetic approach to the sequence-structure
relationship.
Proc Natl Acad Sci USA 1992;89:8721-8725.

\bibitem{Bryngelson1}
Bryngelson JD, Onuchic JN, Socci ND, Wolynes PG.
Funnels, pathways, and the energy landscape of protein-folding --
a synthesis.
Proteins: Struct Funct Genet 1995;21:167-195.

\bibitem{DillChan}
Dill KA, Chan HS.
From Levinthal to pathways to funnels.
Nature Struct Biol 1997;4:10-19.

\bibitem{Tang}
Li H, Tang C, Wingreen NS.
Are protein folds atypical?.
Proc Natl Acad Sci 1998;95:4987-4990.

\bibitem{Vendruscolo}
Vendruscolo M, Maritan A, Banavar JR.
Stability threshold as a selection principle for protein design.
Phys Rev Lett 1997;78:3967-3970.

\bibitem{BauerChan}
Bornberg-Bauer E, Chan HS.
Modeling evolutionary landscapes: mutational stability, topology,
and superfunnels in sequence space.
Proc Natl Acad Sci USA 1999;96:10689-10694.

\bibitem{Michel1} 
Micheletti C, Banavar JR, Maritan A, Seno F.
Protein Structures and optimal folding from a geometrical variational
principle. 
Phys Rev Lett 1999;82:3372-3375.

\bibitem{Michel} 
Maritan A, Micheletti C, Banavar JR.
Role of secondary motifs in fast folding polymers: a dynamical
variational principle.
Phys Rev Lett 2000;84:3009-3012.

\bibitem{Maritan}
Maritan A., Micheletti C., Trovato A., Banavar JR.
Optimal shapes of compact strings.
Nature 2000;406:287-290.

\bibitem{Munoz}
LopezHernandez E., Cronet P., Serrano L., Munoz V.
Folding kinetics of Che Y mutants with enhanced native
alpha-helix propensities.
J Mol Biol 1997;266:610-620.

\bibitem{Plaxco} 
Plaxco KW, Simons KT, Baker D.
Contact order, transition state placement and the refolding rates of 
single domain proteins. 
J Mol Biol 1998;277:985-994.

\bibitem{Ruczinski}
Plaxco KW, Simons KT, Ruczinski I, Baker D.
Topology, stability, sequence, and length: defining the determinants
of two-state protein folding kinetics.
Biochemistry 2000;39:11177-11183.

\bibitem{Chan}
Chan HS.
Matching speed and locality.
Nature 1998;392:761-763.

\bibitem{Goabe} 
Abe H, Go N.
Noninteracting local-structure model of folding and unfolding transition in
globular proteins. II. Application to two-dimensional lattice proteins.
Biopolymers 1981;20:1013-1031.

\bibitem{Hoang} 
Hoang TX, Cieplak M.
Molecular dynamics of folding of secondary structures in Go-like models
of proteins. 
J Chem Phys 2000;112:6851-6862.

\bibitem{Hoang1}
Hoang TX, Cieplak M.
Sequencing of folding events in Go-like proteins. 
J. Chem. Phys. 2001;113:8319-8328.

\bibitem{KimBaker}
Kim DE, Gu H, Baker D.
The sequences of small proteins are not extensively optimized for
rapid folding by natural selection.
Proc Natl Acad Sci 1998;95:4982-4986.

\bibitem{Fersht}
Fersht AR.
Transition state structure as a unifying basis in protein folding mechanisms:
contact order, chain topology, stability and the extended nucleus mechanism.
Proc Natl Acad Sci 2000;97:1525-1529.

\bibitem{Baker}
Baker D.
A surprising simplicity to protein folding.
Nature 2000;405:39-42.

\bibitem{Socci} 
Socci ND, Onuchic JN.
Folding kinetics of protein-like heteropolymers. 
J Chem Phys 1994;101:1519-1528.

\bibitem{Takada} 
Takada S, Wolynes PG. 
Microscopic theory of critical folding nuclei and reconfiguration activation
barriers in folding proteins. 
J Chem Phys 1997;107:9585-9598.

\bibitem{Finkelstein} 
Finkelstein AV, Badredtinov AY.
Rate of protein folding near the point of thermodynamic
equilibrium between the coil and the most stable chain fold.
Fold Des 1997;2:115-121.

\bibitem{Wolynes2} 
Wolynes PG.
Folding funnels and energy landscapes of larger proteins within the
capillarity approximation.
Proc Natl Acad Sci USA 1997;94:6170-6175.

\bibitem{Thirumalai3} 
Thirumalai D. 
From minimal models to real proteins: timescales for protein folding. 
J Physique I 1995;5:1457-1467.
Thirumalai D, Klimov DK.
Deciphering the timescales and mechanisms of protein folding using minimal
off-lattice models.
Curr Opin Struct Biol 1999;9:197-207.

\bibitem{Camacho} 
Camacho CJ.
Entropic barriers, frustration, and order: basic ingredients in
protein folding. 
Phys Rev Lett 1996;77:2324-2327.

\bibitem{Shakhnovich} 
Gutin AM, Abkevich VI, Shakhnovich EI.
Chain length scaling of protein folding time.
Phys Rev Lett 1996;77:5433-5436.

\bibitem{Zhdanov} 
Zhdanov VP.
Folding time of ideal $\beta$ sheets vs. chain length.
Europhys Lett 1998;42:577-581.

\bibitem{Cieplak} 
Cieplak M, Hoang TX, Li MS.
Scaling of folding properties in simple models of proteins.
Phys Rev Lett 1999;83:1684-1687.

\bibitem{Stakada} 
Takada S.
Go-ing for the prediction of protein folding mechanism.
Proc Natl Acad Sci USA 1999;96:11698-11700.

\bibitem{Alm}
Alm E., Baker D.
Matching theory and experiment in protein folding.
Curr Opin Struct Biol 1999;277:189-196.

\bibitem{Kaya}
Kaya H, Chan HS.
Polymer principles of protein calorimetric two-state cooperativity.
Proteins: Struct Funct Genet 2000;40:637-661.

\bibitem{Li} 
Li MS, Cieplak M.
Folding in two-dimensional off-lattice models of proteins. 
Phys Rev E 1999;59:970-976.

\bibitem{Klimov2} 
Klimov DK, Thirumalai D.
Mechanisms and kinetics of $\beta$-hairpin formation.
Proc Natl Acad Sci USA 2000;97:2544-2549.

\bibitem{Zhou} 
Zhou Y, Karplus M.
Interpreting the folding kinetics of helical proteins.
Nature 1999;401:400-403.

\bibitem{Stanley}
Dokholyan NV, Buldyrev SV, Stanley HE, Shakhnovich EI.
Discrete molecular dynamics studies of the folding
of a protein-like model.
Folding Des 1998;3:577-587.

\bibitem{Nymeyer} 
Clementi C, Nymeyer H, Onuchic JN.
Topological and energetic factors: what determine the structural
details of the transition state ensemble and ``on-route'' intermediates
for protein folding? An investigation for small globular proteins.
J Mol Biol 2000;298:937-953.

\bibitem{Veitshans}
Veitshans T, Klimov D, Thirumalai D.
Protein folding kinetics: time scales, pathways and energy landscapes in
terms of sequence-dependent properties.
Folding Des 1997;2:1-22.

\bibitem{PDB} 
Bernstein FC, Koetzle TF, Williams GJB,
Meyer Jr. EF, Brice MD, Rodgers JR, Kennard O, Shimanouchi T,
Tasumi M.
The Protein Data Bank: a computer-based archival file for macromolecular
structures.
J Mol Biol 1977;112:535-542.


\bibitem{Gennes} 
de Gennes PG.
Kinetics of collapse for a flexible coil. 
J Phys Lett 1985;46:L639-L642.

\bibitem{Henkel} 
Cieplak M, Henkel M, Karbowski J, Banavar JR.
Master equation approach to protein folding and kinetic traps.
Phys Rev Lett 1998;80:3654-3657.

\bibitem{Takano}
Takano M, Takahashi T, Nagayama K.
Helix-coil transition and 1/f fluctuation in a polypeptide.
Phys Rev Lett 1998;80:5691-5694.

\end{thebibliography}
\end{document}